\documentclass[prl,aps,superscriptaddress,showpacs]{revtex4}

\usepackage{dcolumn}
\usepackage{bm}
\usepackage{tikz}
\usepackage{epstopdf}
\usepackage[colorlinks=false]{hyperref}
\usepackage{amsmath,amsfonts,amssymb,times,natbib}
\usepackage[standard]{ntheorem}
\def\endproof{\vrule height6pt width6pt depth0pt}

\begin{document}
\title{Beyond Gisin's Theorem and its Applications:
Violation of Local Realism by Two-Party Einstein-Podolsky-Rosen Steering}

%Extension of Gisin's Theorem to mixed states with a two party Einstein-Podolsky-Rosen ``Steering" allows to reduce entanglement complexity in quantum communication protocols}
%\title{Extension of Gisin's Theorem to Mixed States with a Two Party Einstein-Podolsky-Rosen ``Steering" Allows to Reduce Entanglement Complexity in Quantum Communication Protocols}

\author{Jing-Ling Chen\footnote{Correspondence to
J.L.C. (chenjl@nankai.edu.cn).}}
%\email{chenjl@nankai.edu.cn}
 \affiliation{Theoretical Physics Division, Chern Institute of Mathematics, Nankai University,
 Tianjin 300071, People's Republic of China}
 \affiliation{Centre for Quantum Technologies, National University of Singapore,
 3 Science Drive 2, Singapore 117543}

\author{Hong-Yi Su}
 \affiliation{Theoretical Physics Division, Chern Institute of Mathematics, Nankai University,
 Tianjin 300071, People's Republic of China}

\author{Zhen-Peng Xu}
 \affiliation{Theoretical Physics Division, Chern Institute of Mathematics, Nankai University,
 Tianjin 300071, People's Republic of China}

\author{Yu-Chun Wu}
 \affiliation{Key Laboratory of Quantum Information, University of Science and Technology of China, 230026 Hefei, People's Republic of China}
\affiliation{Synergetic Innovation Center of Quantum Information and Quantum Physics, University of Science and Technology of China, 230026 Hefei, Anhui, China}

\author{Chunfeng~Wu}
\affiliation{Pillar of Engineering Product Development, Singapore University of Technology and Design, 8 Somapah Road, Singapore 487372}

\author{Xiang-Jun~Ye}
 \affiliation{Theoretical Physics Division, Chern Institute of Mathematics, Nankai University,
 Tianjin 300071, People's Republic of China}

\author{Marek \.{Z}ukowski}
 \affiliation{Institute of Theoretical Physics and Astrophysics, University of Gda\'{n}sk, PL-80-952 Gda\'{n}sk, Poland}
 \affiliation{Hefei National Laboratory for Physical Sciences at Microscale and Department of Modern Physics, University of Science and Technology of China, 230026 Hefei, China}

\author{L.~C.~Kwek\footnote{Correspondence to
L.C.K. (cqtklc@nus.edu.sg).}}
%\email{cqtklc@nus.edu.sg}
\affiliation{Centre for Quantum Technologies, National
University of Singapore, 3 Science Drive 2, Singapore 117543}
\affiliation{National Institute of Education,1 Nanyang Walk, Singapore 637616}
\affiliation{Institute of Advanced Studies, Nanyang Technological University, 60 Nanyang View,
Singapore 639673}

\date{\today}

\pacs{03.65.Ud,
%Entanglement and quantum nonlocality
%(e.g. EPR paradox, Bell's inequalities, GHZ states, etc.)
03.67.Mn,
%Entanglement production, characterization, and manipulation
42.50.Xa}
%Optical tests of quantum theory

%%%%%%%%%%%%%%%%%%%%%%%%%%%%%%%%%%%%%%%%%%%%%%%%%%%%%%%%%%%%%%%%%%%

\maketitle

%%%%%%%%%%%%%%%%%%%%%%%%%%%%%%%%%%%%%%%%%%%%%%%%%%%%%%%%%%%%%%%%%%%

%\section{Introduction}

%%%%%%%%%%%%%%%%%%%%%%%%%%%%%%%%%%%%%%%%%%%%%%%%%%%%%%%%%%%%%%%%%%%

\textbf{We demonstrate here that for a given mixed multi-qubit state if there are at least two observers for whom mutual Einstein-Podolsky-Rosen steering is possible, i.e. each observer is able to steer the other qubits into two different pure states by spontaneous collapses due to von Neumann type measurements on his/her qubit, then nonexistence of local realistic models is fully equivalent to quantum entanglement (this is not so without this condition). This result leads to an enhanced version of Gisin's theorem (originally: all pure entangled states violate local realism). Local realism is violated by all mixed states with the above steering property. The new class of states allows one e.g. to perform three party secret sharing with just pairs of entangled qubits, instead of three qubit entanglements (which are currently available with low fidelity). This significantly increases the feasibility of having high performance versions of such protocols. Finally, we discuss some possible applications.}

Quantum mechanical correlations do not admit local realistic models.  This pivotal result  concerns the  foundations of quantum mechanics (QM) and has many applications in  quantum
information theory.  The states exhibiting quantum correlations can be grouped using the following hierarchy ~\cite{WJD07}: those that are entangled, those that allow for  Einstein-Podolsky-Rosen (EPR) ``steering",  and those that violate local realism (LR)~\cite{Bell}.

According to Edwin Schr\"odinger ~\cite{Schrodinger35}, quantum entanglement is ``the
characteristic trait of quantum mechanics" distinguishing it from any classical theory. This feature of quantum states is a highly useful resource in many fascinating applications of quantum information,
such as teleportation~\cite{telep1,telep2}, dense coding, communication
protocols, and computation~\cite{Nielsen,PAN}.  Moreover, states that violate
local realism have gained ubiquitous applications in different quantum
information tasks, such as quantum key distribution~\cite{crypt},
communication complexity~\cite{Brukner}, quantum information processing
and  random number
generation~\cite{Random10}.

In Schr\"odinger's reply % \cite{SchrodingerReply}
 to the EPR paper~\cite{EPR}, he made a fine distinction
between entangled states and states shared between two parties that are amenable for steering, i.e. in which
the action of one party can affect the reduced state of the other party through the an appropriate
choice of measurements.
Also, in contrast to quantum entanglement which has received widespread interest due to its usefulness as a resource
for tasks in quantum information processing, there have been relatively fewer developments in the notion of ``steerable"
states.  In 2007, Wiseman {\it et al.}~\cite{WJD07} revisited the issue and reformulate the idea of ``steerable" state
from a quantum information perspective.
Since then,  several interesting studies on  EPR steering have appeared both in theory
~\cite{SU,QKD}  and in experiments ~\cite{NP2010,NC,PRX,NJP, AS12}. Recently, some of us have demonstrated an
all-versus-nothing proof of EPR steering~\cite{AVN}.

Quantum steering in  a bipartite scenario essentially describes the ability of one party, say Alice, to
prepare the other party's (Bob's) systems in different ensembles of quantum states
by measuring her own particle (under more than two different settings).
Naturally, Alice has no control over the actual result of her actions,
but she is able to control over the set of projected states
on Bob's side  with her measurement for a given setting.
If Bob does not
trust Alice, Alice's manipulation of her system  appears as
a black box described by some local hidden variable (LHV) theory. This could imply  a local hidden state (LHS) description of her actions: that is that she just simply sends some states to Bob according to some probability distribution. Quantum steering  means that such a description is impossible and that Alice must use quantum measurements on her system to prepare states on Bob's side.
%Note that LHS models mix the rules of quantum mechanics (Bob's system) with classical theories  (Alice's box).

As pointed by  Wiseman \emph{et al.}~\cite{WJD07}, there is a hierarchical structure.
For a given state, quantum steering is strictly implied by  the violation of local realism. Simply put,
steering excludes the possibility of local ``hidden" state
 models of correlations,  in which a quantum
mechanical model is applied to only one of the systems, while the other one is described using a local hidden variable model. LHS model allows for a full LHV model for both (all) systems. Of course separable (mixed) states can be thought of as  probabilistic distributions of hidden states for each party, and thus states with steering are a proper subset of entangled states, endowed with an additional potentially useful property.

%
%\begin{figure}[tbp]
%\includegraphics[width=70mm]{newfig1.eps}\\
%\caption{The hierarchic structure of  quantum correlations. Quantum states violating local realism form a subset
%of EPR steerable states which, in turn, form a subset of entangled
%states.  }\label{fig}
%\end{figure}

The close connection between quantum entanglement and violation of local realism~\cite{Werner} can be traced
back to Gisin's  work,  Ref. \cite{Gisin}, in which he presented a
theorem stating that any pure entangled state of two qubits
violates a Bell-like inequality. The result was generalized in Refs.~\cite{Popescu-Rohrlich,GISIN-PERES}.
% Gisin discussed this simple result with John Bell and it appeared that Bell himself did not know about this property of pure states ~\cite{Gisin}.
Gisin's theorem for three qubits was shown numerically by Chen
\emph{et al.}~\cite{Chen} and analytically by Choudhary \emph{et
al.}~\cite{Choudhary}. In 2012, Yu \emph{et al.}~\cite{Yu} provided a
complete proof of Gisin's theorem for all entangled pure states.
Within the hierarchical picture of the states exhibiting quantum correlations, Gisin's theorem
says that violation
 of local realism and quantum
entanglement are equivalent for all \emph{pure} states.

Gisin's theorem  applies only to
pure states. The aim of this paper is to develop an ``enhanced" version of
Gisin's theorem that can apply also  to some class of mixed states.
%EPR steering is something in between just  quantum entanglement
%and the stronger property:  violation of some Bell inequalities. THIS WAS SAID EARLIER
We shall show that some form of  EPR steering allows the Gisin's
theorem to be applicable to a wider range of entangled states than just pure ones. As pure entangled states always allow steering, we have a direct broadening of the realm of validity of Gisin's theorem.  Our
result also provides a rigorous criterion for marking the borders between
quantum entanglement, EPR steering and violation of local realism. This is a
nontrivial problem since it is not easy to reduce a superset to a subset by imposing
extra constraints.

Our result serves as a contribution to recent  studies of general nonlocal
theories which satisfy the non-signaling principle~\cite{PR,nonlocal}
(or its extensions~\cite{Fritz}), asking  the  question under what conditions
an entangled state does not violate local realism~\cite{ent-non}. Other links are with studies of the
role of quantum contextuality in violation of local realism like e.g. Ref. ~\cite{adan}.
In \textbf{Methods},
%Sec. \ref{enhanced},
we prove two theorems regarding the EPR steerability. We also note that the original Gisin's theorem is a special case of Theorem 1. We note that our criterion for steerability of quantum states is useful in some applications.  %In Sec. \ref{sec:thirdman},
We then describe how our criterion for steerability of quantum states could be applied to the Third Man cryptographic protocol and %in Sec. \ref{sec:app2},
we show how Theorem 2 %in Sec. \ref{enhanced}
can serve as a valuable resource in a quantum certificate authorization protocol.

\vspace{5mm}
\noindent\textbf{Results}

%\section{Enhanced Gisin's theorem} \label{enhanced}

\noindent\textbf{Enhanced Gisin's theorem.}
Gisin's original work
only applies to two qubits in a pure state. In the spirit of this, let us also start with the two-qubit case, for which we
have the following theorem:

\begin{theorem}
For a two-qubit entangled state, $\varrho$, shared between Alice and Bob,
if any party can steer the other party into two different pure states
by performing some orthogonal projective measurements
$\{\mathcal{P}_{\vec{n}}^0, \mathcal{P}_{\vec{n}}^1\}$ on her/his
qubit, then the state violates local realism.
\end{theorem}

\emph{Proof.} Take the spectral decomposition of  two-qubit density matrix:
 $\varrho=\sum_{i=1}^4 \nu_i
|\Psi_i\rangle\langle\Psi_i|$,
%\begin{eqnarray}
% \varrho=\sum_{i=1}^4 \nu_i |\Psi_i\rangle\langle\Psi_i|.
%\end{eqnarray}
the positive $\nu_i$  add up to one. For  $|\Psi_i\rangle$, if Alice performs the
orthogonal projective measurements on  on her qubit given by $\mathcal{P}_{\vec{n}}^0=|+\vec
n\rangle\langle +\vec n|, \mathcal{P}_{\vec{n}}^1=|-\vec
n\rangle\langle -\vec n|$ along the $\vec n$-direction and
is able in this way to steer Bob's qubit into two different states $|\chi_1\rangle$
and $|\chi_2\rangle$, then the state $|\Psi_i\rangle$  must be in the
form
\begin{eqnarray}
|\Psi_i\rangle= F_i |+\vec n\rangle|\chi_1\rangle+ \sqrt{1-F_i^2} \;e^{i\tau} |-\vec
n\rangle|\chi_2\rangle,
\end{eqnarray}
with real $F_i$ satisfying  $0<F_i<1$.
However, in the formula we have only two mutually orthogonal vectors $|+\vec
n\rangle|\chi_1\rangle$, $|-\vec n\rangle|\chi_2\rangle$. This means that we can have only the
two independent $|\Psi_i\rangle$ with the same steering property. Hence, the rank of $\varrho$
is at most 2:
\begin{eqnarray}
 \varrho=\nu_1 |\Psi_1\rangle\langle\Psi_1|+\nu_2
 |\Psi_2\rangle\langle\Psi_2|. \label{twoqubit}
\end{eqnarray}
  If also Bob with measurements on  his
qubit  can  steer Alice's qubit into two different pure
states, this additionally constrains the form of the two pure states to
\begin{subequations}\label{phi12}
\begin{eqnarray}
&&|\Psi_1\rangle=\cos\frac{\zeta}{2}|00\rangle+\sin\frac{\zeta}{2}
e^{i\tau}|11\rangle,\\
&&|\Psi_2\rangle=\sin\frac{\zeta}{2}|00\rangle-\cos\frac{\zeta}{2}
e^{i\tau}|11\rangle.
\end{eqnarray}
\end{subequations}
For a derivation see \textbf{Methods}.

%and $\nu_1,\nu_2\geq 0$, $\nu_1+\nu_2=1$, $\zeta$ and $\tau$ are
%real-valued.

A mixture of the such states, with $\nu_1\neq\nu_2$, always violates  the Clauser-Horne-Shimony-Holt
(CHSH) inequality.
In quantum mechanics the correlation
function is computed using $Q_{ij}=
Q_{\vec{n}_{A_i}\vec{m}_{B_j}}
={\rm tr}(\varrho
\vec{\sigma}_{\vec{n}_{A_i}}\otimes\vec{\sigma}_{\vec{m}_{B_j}})$,
where $\vec{\sigma}_{\vec{n}_{A_i}}=\vec{n}_{A_i}\cdot\vec{\sigma}$, and $\vec{\sigma}$ is the Pauli matrix vector,
whereas  $\vec{n}_{A_i}$ is the $i$-th measuring direction of Alice.
We shall use the spherical coordinates, so that
$\vec{n}_{A_i}=(\sin\theta_{A_i}\cos\phi_{A_i},\sin\theta_{A_i}\sin\phi_{A_i},\cos\theta_{A_i})$;
similarly for Bob.
%Let $\nu_1=(1+V)/2$ and $\nu_2=(1-V)/2$, then
%(\ref{twoqubit}) becomes $\varrho=V|\Psi_1\rangle\langle
%\Psi_1|+(1-V)\openone_{\rm c}/2$, with $V\in [0,1]$ the visibility
%and $\openone_{\rm c}=|00\rangle\langle00|+|11\rangle\langle11|$ the
%so-called colored noise.
The local realistic CHSH constraint is that  $\mathcal
{I}_{\rm CHSH}=\frac{1}{2}(Q_{11}+Q_{12}+Q_{21}-Q_{22})\leq1$.
By putting
$\phi_{A_1}=\phi_{B_1}=\phi_{B_2}=\theta_{A_1}=0,\;\;\theta_{A_2}=\pi/2,\;\;\theta_{B_2}=-\theta_{B_1},\;\;\phi_{A_2}=\tau$,
we get $ \mathcal {I}^{\rm max}_{\rm CHSH}=\sqrt{1+\mathcal
{C}^2}$, where $\mathcal {C}=|V\sin \zeta|$ is the degree of
entanglement, and  $V=\nu_1-\nu_2$. Except at
$\zeta=0,\pi$ or $V=0$, $\mathcal {C}$ is always nonvanishing, and the
CHSH inequality is  violated. \hfill\endproof

Note that, the original Gisin's theorem is a special case of
Theorem 1.  For example, take $\nu_1=1$.  The state $|\Psi_1\rangle$, for $\zeta\neq0$ or $\pi,$ is a
 two-qubit pure state in its Schmidt decomposition.

%We can also restrict $\nu_1$ within the domain
%$[\frac{1}{2},1]$, i.e., $V\in [0,1]$ (note that the state is
%separable when $\nu_1=1/2$), since a state $\varrho$ with other
%values of $\nu_1$ can be locally transformed to this by
%$\openone\otimes\sigma_z$ and the substitution $\zeta\rightarrow
%\pi-\zeta$. THIS IS IRRELEVANT,I THINK

%To see it clearly, let us consider the two-qubit
%pure state shared by Alice and Bob in the Schmidt form
%\begin{eqnarray}
%|\Psi\rangle=\cos\frac{\zeta}{2}|00\rangle+\sin\frac{\zeta}{2}|11\rangle.
%\end{eqnarray}
%Suppose Alice measures her qubit by the projectors
%$\mathcal{P}_{\hat{n}}^a=\frac{1}{2}(\openone+(-1)^a\hat{n}\cdot\vec{\sigma})$,
%then it is easy to find that Bob's qubit collapses to two different
%pure states, unless $\Psi$ is separable.

For the general case of $N$ qubits, we have the following theorem:

\begin{theorem}
For an $N$-qubit entangled state shared by $N$ observers
$\mathcal{O}_1, \mathcal{O}_2,\cdots,  \mathcal{O}_N$, if there exists
at least two observers, each with the ability to steer the remaining $N-1$
qubits into two different pure states by performing some orthogonal
projective measurements on their qubits, then the state violates
local realism.
\end{theorem}

Without any loss of generality, we assume that the first two
observers have the ability to steer the remaining $N-1$ qubits. As
such, the state $\varrho$ is either (a) an arbitrary $N$-qubit
entangled pure state or (b) a rank-2 density matrix as in
(\ref{twoqubit}) with
\begin{subequations}\label{Nqubit}
\begin{eqnarray}
|\Psi_1\rangle &=& \cos \frac{\zeta}{2} |0_1 0_2\rangle \otimes |\chi_1\rangle + \sin \frac{\zeta}{2}e^{i\tau} |1_1 1_2\rangle \otimes |\chi_2\rangle, \\
|\Psi_2\rangle &=& \sin \frac{\zeta}{2} |0_1 0_2\rangle \otimes
|\chi_1\rangle - \cos \frac{\zeta}{2} e^{i\tau}|1_1 1_2\rangle
\otimes |\chi_2\rangle,
\end{eqnarray}
\end{subequations} where $|\chi_1\rangle,|\chi_2\rangle$ are pure
states for $\mathcal{O}_3,\cdots,\mathcal{O}_N$, and the relative
phase $\tau$ can always be taken as zero.

The violation of local realism for Case (a) has been shown analytically
in~\cite{Choudhary,Yu} with the use  a generalized Hardy
(see e.g. inequality~\cite{Cereceda1,Cereceda2,Cereceda3}). To
prove Case (b),  it is convenient to consider first $N=3$, before moving to $N\geq 4$. (See \textbf{Methods} for the rigorous proof.)

%The sketch of the proof is that when
%$k\in\{3,\cdots,N-1\}$ we utilize the Hardy inequality, otherwise we
%devise a novel family of Bell inequalities, as will be shown later.

%Then Charlie randomly asks Alice and Bob to measure along one of two optional directions: $\{\hat{x}_A\otimes\hat{x}_B, \hat{z}_A\otimes\hat{z}_B\}$. To ensure the states are faithfully distributed, Charlie randomly picks part of those measurements and requires Alice and Bob to broadcast their results through the public channel. If the state shared by Alice and Bob is reliable,  they should get the same results (cf. Eqs.~(\ref{Nqubit})) for any measurement  along $\hat{z}$. Thus, if Alice's and Bob's results are both $0$, then Charlie's qubit is along $|\chi_1\rangle$, or if their results are both $1$, then Charlie's qubit is along $|\chi_2\rangle$. Finally, they examine their results: (a) whether Alice (Bob) can steer Bob (Alice) and Charlie into two pure states when measuring along $\hat{z}$, and (b) whether $V  \sin \zeta\neq0$~\cite{aspect2}. Result (a) ensures that both Alice and Bob can steer the remaining parties into two pure states; and if (a) is true, result (b) certifies that the state $\varrho$ is entangled. Hence, if both (a) and (b) are satisfied, then according to \textbf{Theorem 2} the state $\varrho$ violates local realism, and the protocol is secure. Note that the secret keys are obtained from the unbroadcast part of results when Alice and Bob measure along $\hat{z}$.

%\section{Application 1: The Third Man cryptography}\label{sec:thirdman}
\noindent\textbf{Application 1: The Third Man cryptography.}
We have extended the class of states for which  Gisin's Theorem holds. But are these new states endowed with properties that can be put to use in quantum information tasks, do they form a kind of new resource? Below we shall show that in specific cases they can reduce the number  of entangled particles needed to perform a task. In our example for three partners  this is The Third Man cryptography \cite{ZUK1,ZUK2} or equivalently secret sharing \cite{BUZEK}.  The example is extendible to more partners.

Imagine that Charlie is sending the states of Theorem 1 to Alice and Bob. He randomly chooses whether to send $|\Psi_1\rangle$ or  $|\Psi_2\rangle$, with probabilities $\nu_1$ and $\nu_2$, which are different but quite similar. To simplify the example, assume that both pure states are maximally entangled, that is $\zeta=\pi/2$. Alice and Bob are asked to perform randomly chosen measurements using local bases $\vec{x}$ and $\vec{y}$, and also in some auxiliary bases in the $xy$ planes, usually needed in the Ekert91 protocol \cite{crypt}.
But the trouble is that while one of the pure states gives perfectly correlated results when both measured in  $\vec{x}$  direction, and anti-correlated results for measurements in directions $\vec{y}$, the other one gives perfect anti-correlations for $\vec{x}$ measurements and perfect correlations for $\vec{y}$'s (it is irrelevant for the argument for which of the states this is so). Thus only if Charlie sends them information about which state he sent in the given run, they can unscramble the key, out of their measurements in the identical local bases. Thus Charlie holds a key to their key. Without additional information provided by him Alice and Bob cannot form a usable key. This is the Third Man cryptography.

Of course, if Charlie sends information on the states to just Alice, she can transform her string of data into one which is a copy of the string of Bob, and she can also build a working key (this is a secret sharing version of the same protocol). Of course, all this is done under the standard Ekert91 protocol, Alice and Bob exchange information on measurements bases without revealing the results, etc. Note, that for high, say $\nu_1$, with classical error correction methods Alice and Bob could be able to extract a key, but it will be short, with respect to the numbers of runs (copies of the state sent by Charlie). Only with the help of Charlie it can be of maximal possible length (half of the numbers of runs, if the protocol runs perfectly).

Previous versions of such protocol, see \cite{ZUK1,ZUK2} or \cite{BUZEK}, required three qubit entangled GHZ states. The reader may quickly judge how difficult the step is from two-particle entanglement distribution to three-particle entanglement distribution by consulting the review \cite{PAN}. Also the fidelity of photonic three-particle entangled states is currently still typically below 90\%, while in the two-particle case it can be now well over 99\%. Low fidelity leads to errors in the key distribution. Thus the scheme with steerable states has a clear advantage over the the original one with GHZ states.

There is one more advantage. By making measurements in the $\vec{z}$ directions, which are always perfectly correlated, Alice and Bob can check whether the mixed state received by them from  Charlie ($\varrho$, that is the state before he reveals in  which runs were  $|\Psi_1\rangle$ or  $|\Psi_2\rangle$ ) is indeed entangled (as it violates the CHSH inequality). Thus they can have an independent quality check of Charlie's distribution methods. This is impossible under GHZ stated based protocols.

Note that the protocol settings for Alice and Bob are $\vec{x}$ and $\vec{y}$. With settings in $\vec{x}$ and $\vec{y}$ Alice and Bob cannot violate the CHSH inequality with $\varrho$ for which $\nu_1$ and $\nu_2$ are close. Using the Horodecki criterion for violations of the CHSH inequality  \cite{HOR-CRIT}, $ \sum_{i.j=\vec{x},\vec{y}}Q_{ij}^2>1$,   given here in the form derived in  \cite{ZB}, one can easily establish that the threshold difference is $|\nu_1-\nu_2|= {1}/{\sqrt{2}}$. The correlations in these directions are very weak, thus they are indeed unable to extract a key with such measurements, and the Ekert91 protocol. Without Charlie's help they are helpless. Of course, the presented scheme can be modified in many ways.

%\section{Application 2: The quantum certificate authorization protocol}\label{sec:app2}
\noindent\textbf{Application 2: The quantum certificate authorization protocol.}
The state $\varrho$ we discussed in \textbf{Theorem 2}
%in the main text
can serve as a valuable resource in quantum cryptography~\cite{scarani}. A good example is an application to quantum certificate authorization involving three parties, say, Alice, Bob and Charlie, against a lurking eavesdropper, Eve. Suppose Alice needs to send some private information to Bob through internet, and yet she is not sure if the receiving party is Bob. So Alice goes to the certificate authority, Charlie, who is trusted by Alice and capable of certifying Bob's identity. With Charlie's help, Alice is then able to share secret keys with Bob at a distance.

There are two goals to be achieved here. Specifically, Alice's private information (i) should be received by the true Bob (and not someone else who claims to be Bob) and (ii) should not be intercepted by Eve. This task can be classically realized through digital signatures and public-private keys. Given $\varrho$, we see that we are able to present a quantum analogue of this protocol (see also~\cite{qca1,qca2}). As shown in Fig. 1, upon Alice's request for identification of Bob, Charlie produces an ensemble of three-qubit states $\varrho$'s and distributes the first qubit to Alice, the second qubit to Bob, and keeps the third qubit. Alice and Bob measure their qubits randomly along one of two directions: $\vec{z}$ or $\vec{x}$ (i.e., projections to $|0\rangle\langle0|,|1\rangle\langle1|$ or $|+\rangle\langle+|,|-\rangle\langle-|$); similarly Charlie also measures his qubit randomly by a projection to $|\chi_1\rangle\langle\chi_1|$ or $|\chi_2\rangle\langle\chi_2|$. Such a joint measurement can be repeated for, say, $N$ times, where $N$ is large enough. If the state shared by Alice and Bob is reliable,  they should get the same results (cf. Eqs.~(4) in the main text) for any measurement  along $\vec{z}$.

In order to detect a possible Eve, Charlie randomly picks $m$ runs from $N$ as a subset and requests Alice and Bob to broadcast through a public channel their directions and results for these $m$ runs. With their data at hand Charlie starts to do analysis in three steps: (i) keep joint measuring results for which Alice and Bob measured along the same direction, i.e., $\vec{z}_A\otimes\vec{z}_B$ and $\vec{x}_A\otimes\vec{x}_B$, and discard those for $\vec{x}_A\otimes\vec{z}_B$ and $\vec{z}_A\otimes\vec{x}_B$; (ii) keep joint measuring results for which Charlie measured along $|\chi_1\rangle$ ($|\chi_2\rangle$) when Alice and Bob obtained both ``0" (``1") along $\vec{z}$; (iii) verify (a) whether Alice (Bob) can steer Bob (Alice) and Charlie into two pure states when measuring along $\vec{z}$, and (b) whether $V  \sin \zeta\neq0$.

Here (a) can be verified by examining whether Charlie's probability
of obtaining $|\chi_1\rangle$ ($|\chi_2\rangle$) is always unity
when Alice and Bob obtained both ``0" (``1") along $\vec{z}$. For
(b), the three-qubit state $\varrho$ is entangled iff $V \sin
\zeta\neq0$. To realize the quantum certificate authorization
protocol, Charlie can produce $\varrho$ with nonorthogonal
$|\chi_1\rangle$ and $|\chi_2\rangle$. The coincidence probability
that Alice and Bob get the same result can be obtained as
$\frac{1}{2} (1+V  \cos \phi  \sin \zeta )$ when Alice and Bob
measure along $\vec{x}$. To check whether $\varrho$ is entangled is
equivalent to check whether the coincidence probability is not equal
to $1/2$. In other words, result (a) ensures that both Alice and Bob can steer the remaining parties into two pure states; and if (a) is true, result (b) certifies that the state $\varrho$ is entangled. Hence, if both (a) and (b) are fulfilled, then according to \textbf{Theorem 2} the state $\varrho$ violates local realism, and the protocol is secure. Note that the secret keys are obtained from the unbroadcast part of results when Alice and Bob measure along $\vec{z}$.

%\section{Conclusions}

\vspace{5mm}
\noindent\textbf{Discussion}

%Our ``enhanced" Gisin's theorem
%involving  the EPR steering strengthens the connection between quantum entanglement  and violations of
%local realism. The EPR steering sheds a new light here, this is mainly due to the hierarchical structure of quantum entanglement.
%THIS VAGUE< AND ESSENTIALLY THE SAME AS BELOW
\noindent Our result sheds new light on the relation between entanglement of mixed states, and LHV models for correlations.
It pinpoints a precisely defined class of states,  which is strictly  larger than all pure entangled states, and which has the property that  just by the fact that a state  belongs to the class, one knows that it violates local realism.  The property is as follows:  for an $N$-qubit state, there exist a pair of observers such that each of them singlehandedly can steer the remaining $N-1$ qubits of  the other observers into two different pure states. Our results are for qubit systems. A generalization to more complicated ones is
still an open question, and under investigation. Mixed states covered by the enhanced Gisin theorem,  due to their specific properties, may allow new quantum protocols which allow to reduce the complication of entangled states involved in them (as in our example, we need, e.g., two-particle entangled states to have secret sharing between three parties). Other protocols, such as quantum certificate authorization, are also possible.
%(see Appendix 3).

\vskip2ex

\vspace{5mm}
\noindent\textbf{Methods}

%\emph{Appendix: Proof of Eq.~(\ref{phi12}).---}
\noindent \textbf{Proof of Eq.~(\ref{phi12}).} Without loss of generality, if Alice measures her
qubit in the $z$-direction and if she steers Bob's qubit into two different
pure states, then for the states (\ref{twoqubit}) we have
\begin{eqnarray}
|\Psi_1\rangle=\cos\frac{\zeta}{2}|00\rangle+\sin\frac{\zeta}{2}(\cos\beta
e^{i\tau_1}|10\rangle+\sin\beta e^{i\tau_2}|11\rangle),\nonumber\\
|\Psi_2\rangle=\sin\frac{\zeta}{2}|00\rangle-\cos\frac{\zeta}{2}(\cos\beta
e^{i\tau_1}|10\rangle+\sin\beta
e^{i\tau_2}|11\rangle).\nonumber
\end{eqnarray}
However,  if Bob measures his qubit along some specific
direction, also Alice should be getting one of two steered states. We can denote Bob's
projectors by $\{|\xi\rangle\langle\xi|,|\xi'\rangle\langle\xi'|\}.$
The respective  qubit basis states can be put as
\begin{eqnarray}
&&|\xi\rangle=a|0\rangle+b|1\rangle, \;\;\;
|\xi'\rangle=b^*|0\rangle-a|1\rangle
\end{eqnarray}
with $a$ real-valued, $b=\sqrt{1-a^2}e^{i\gamma}$.
If the state of a qubit is pure, then the determinant of its density matrix is zero.
 If
Bob measures his qubit of $\varrho$, given by (\ref{twoqubit}), with the above projectors, the
determinants of the steered states of Alice's qubit are
\begin{eqnarray}
&&{\rm Det}[\rho_{|\xi\rangle}]=a^2 \nu_1\nu_2
|a\cos\beta+b\sin\beta
e^{i(\tau_1-\tau_2)}|^2,\label{1}\\
&&{\rm Det}[\rho_{|\xi'\rangle}]=|b|^2 \nu_1\nu_2
|b\cos\beta-a\sin\beta e^{i(\tau_1-\tau_2)}|^2.\label{2}
\end{eqnarray}
We must find conditions for both of them to vanish. For the only interesting case of both $\nu_i$ greater than zero, we  have to consider two  situations.
 (i) Suppose
$a=0$, then it is enough to have $|\cos\beta|=0$. (ii)
Suppose  $a\neq 0$, then one has the
conditions $|a\cos\beta+b\sin\beta e^{i(\tau_1-\tau_2)}|=0$ and
$|b|^2|b\cos\beta-a\sin\beta e^{i(\tau_1-\tau_2)}|=0$. The only
solution is $a=1$ (i.e., $|b|=0$) and $|\cos\beta|=0$. This directly
leads to (\ref{phi12}). For the general $N$ qubits, one  has
a straight ahead extension of this reasoning leading to Eq.~(\ref{Nqubit}).

%\subsection{Proof of Theorem 2 for $N=3$}

\noindent\textbf{Proof of Theorem 2 for $N=3$.}
Take three observers Alice, Bob and Charlie. Assume that  Alice and Bob have the ability to steer the
remaining two qubits. Due to local unitary, one can always work with
$|\chi_1\rangle_{C}=|0\rangle$,
$|\chi_2\rangle_C=\cos\phi|0\rangle+\sin\phi|1\rangle$, and $\sin
\frac{\zeta}{2}$, $\cos \frac{\zeta}{2}$, $\sin\phi$,
$\cos\phi\geq0$.

%To sketch the proof, we observe that when $\cos\phi\neq0$, we can use the
%Hardy inequality, otherwise we devise alternative Bell
%inequalities as shown later.

(i) Let us first consider the case where $\cos\phi\neq0$. The
three-qubit Hardy inequality is given by
\begin{eqnarray}
% \nonumber to remove numbering (before each equation)
\mathcal{I}_{\rm
Hardy}=p(000|111)-p(111|222)-p(000|112)\;\;\;\;\nonumber\\-p(000|121)
-p(000|211)\leq 0,
\end{eqnarray}
where $p(abc|ijk)$ denotes the probability $p(A_i=a,B_j=b,C_k=c)$.
%For three qubits, if Alice can AVN steer the remaining observers's
%(Bob and Charlie) state and Bob can also AVN steer the remaining
%observers' (Alice and Charlie) state, then the allowed state must be
%in the following form (up to local unitary transformations)
%\begin{eqnarray}
%% \nonumber to remove numbering (before each equation)
%  \rho = V |\psi_1\rangle \langle \psi_1| + (1-V)|\psi_2\rangle \langle \psi_2|,
%\end{eqnarray}
%where
%\begin{eqnarray}
%% \nonumber to remove numbering (before each equation)
% && |\psi_1\rangle = \cos\theta|000\rangle+\sin\theta\cos\phi|110\rangle+\sin\theta\sin\phi|111\rangle,\\
%  &&  |\psi_2\rangle = -\sin\theta|000\rangle+\cos\theta\cos\phi|110\rangle+\cos\theta\sin\phi|111\rangle.
%\end{eqnarray}
%We have some observations:
%
%Observation 1: We can focus ourselves on the special case where
%$\cos\theta,\sin\theta,\cos\phi,\sin\phi$ are all positive. Other
%cases can be locally transformed to this by
%$\openone\otimes\openone\otimes\sigma_z,\openone\otimes\sigma_z\otimes\openone,\openone\otimes\sigma_z\otimes\sigma_z$.
%We can restrict $\nu_1$ within the domain $[\frac{1}{2},1]$, i.e.,
%$V\in [0,1]$ (note that the state is separable when $\nu_1=1/2$),
%since a state $\varrho$ with other values of $\nu_1$ can be locally
%transformed to this by $\openone\otimes\sigma_z\otimes\openone$ and
%the substitution $\zeta\rightarrow \pi-\zeta$.
Let the settings be $ \theta_{A_1}=\theta_{B_1}=\theta_{C_1}=0,
\phi_{A_1}=\phi_{B_1}=\phi_{C_1}=0,
\theta_{C_2}=\pi,\phi_{C_2}=0,\phi_{A_2}=\phi_{B_2}=\frac{\pi}{2},\theta_{B_2}=\theta_{A_2}$,
the quantum prediction then becomes
\begin{eqnarray}
% \nonumber to remove numbering (before each equation)
\mathcal{I}_{\rm Hardy}^{\rm
QM}=\frac{\cos^4\frac{\theta_{A_2}}{2}}{2}\biggr(
2V\sin\zeta\cos\phi\tan^2\frac{\theta_{A_2}}{2}\nonumber\\-(1+\cos^2\phi+V\cos\zeta\sin^2\phi)
 \biggr).
\end{eqnarray}
%Note that the first and the second terms in the bracket are both
%positive. For simplicity, we carefully take the value of
%$\theta_{A_2}$ such that the second term is equal to half the first
%term.
By taking
\begin{eqnarray}
% \nonumber to remove numbering (before each equation)
\tan^2\frac{\theta_{A_2}}{2}=\frac{(1+\cos^2\phi+V\cos\zeta\sin^2\phi)}{V\sin\zeta\cos\phi},\label{ss}
\end{eqnarray}
we finally have
\begin{eqnarray}
% \nonumber to remove numbering (before each equation)
\mathcal{I}_{\rm Hardy}^{\rm
QM}=\frac{\cos^4\frac{\theta_{A_2}}{2}}{2}\left(
1+\cos^2\phi+V\cos\zeta\sin^2\phi \right)>0.\nonumber
\end{eqnarray}

(ii)  If $\cos\phi=0$, we employ the following Bell inequality:
\begin{eqnarray}
% \nonumber to remove numbering (before each equation)
\mathcal{I}_{3}=\frac{1}{4}\biggr(
Q_{111}+Q_{121}+Q_{211}+Q_{221}+Q_{110}\nonumber\\+Q_{120}+Q_{210}-3Q_{220}\biggr)\leq1,
\end{eqnarray}
where $Q_{ijk}$ are the qubit correlation functions defined in analogy to the two qubit ones (see above). The index $0$
indicates that  measurement  performed on the corresponding qubit do not enter the function, and thus we have a two-qubit correlation function. By
taking settings as
$\phi_{A_1}=\phi_{B_1}=\phi_{C_1}=\phi_{A_2}=\phi_{B_2}=\phi_{C_2}=0$,
$\theta_{B_1}=\theta_{C_1}=\frac{\pi}{2}$,
$\theta_{B_2}=\theta_{C_2}=\pi$, and $\theta_{A_2}=0$, all
correlation functions vanish except the following ones:
$Q_{111}=V\sin\zeta\sin\theta_{A_1}$, $Q_{120}=-\cos\theta_{A_1}$,
and $Q_{220}=-1$. The quantum prediction becomes
\begin{eqnarray}
\mathcal {I}_3^{\rm QM}
&=&1+\frac{\cos^2\frac{\theta_{A_1}}{2}}{2}(V\sin\zeta\tan\frac{\theta_{A_1}}{2}-1).\nonumber
\end{eqnarray}
Clearly, we have $\mathcal {I}_3^{\rm QM}>1$ when
$\tan\frac{\theta_{A_1}}{2}>\frac{1}{V\sin\zeta}$.  \hfill\endproof

%\subsection{Proof of Theorem 2 for cases $N\geq4$}

\noindent\textbf{Proof of Theorem 2 for Cases $N\geq4$.}
For $N\geq4$, the states
$|\chi_1\rangle,|\chi_2\rangle$ can be entangled. The proof can be split into two cases: (i) separable $|\chi_1\rangle,|\chi_2\rangle$, and (ii) at least one of them being entangled. To demonstrate violations of local realism, we employ an $N$-qubit Hardy inequality, together with some Bell inequalities devised by us particularly for the present paper.
% (see Appendix 2 for details).

%For $N\geq4$, in contrast to $N=3$, the states
%$|\chi_1\rangle,|\chi_2\rangle$ are entangled states.
The
$N$-qubit Hardy inequality reads
\begin{eqnarray}
% \nonumber to remove numbering (before each equation)
\mathcal{I}^N_{\rm
Hardy}=p(00\cdots0|11\cdots1)-p(11\cdots1|22\cdots2)\nonumber\\-\sum
p(00\cdots0|{\rm Perm}[11\cdots12])\leq0.
\end{eqnarray}
%will play an essential role in the proof.
Here ${\rm
Perm}[11\cdots12]$ is any permutation of indices between parties,
and the summation is taken over all such permutations that are
possible. We see that we have the following cases:

(i) One, or both, of $|\chi_1\rangle,|\chi_2\rangle$ is entangled.
If $|\chi_1\rangle$ is entangled, one can use the following
Bell inequality
\begin{eqnarray}
% \nonumber to remove numbering (before each equation)
  p(0_1 0_2|1_1 1_2)\otimes \mathcal{I}^{N-2}_{\rm Hardy} \equiv p(0000\cdots0|1111\cdots1)\;\;\;\;\nonumber\\-p(0011\cdots1|1122\cdots2)\;\;\;\;\nonumber\\-\sum
p(0000\cdots0|11{\rm Perm}[11\cdots12])\leq0
\end{eqnarray}
to detect violation of local realism. Here $\mathcal{I}^{N-2}_{\rm Hardy}$
is the $(N-2)$-qubit Hardy inequality for
$\mathcal{O}_3,\mathcal{O}_4,\cdots,\mathcal{O}_N$. The
validity of this inequality to test the violation of
local realism of $\varrho$ relies on the fact that when $\mathcal{O}_1$ and
$\mathcal{O}_2$ measure their qubits respectively along the
$z$-direction, these measurements are equivalent to a test for the violation of local realism of
$|\chi_1\rangle$ using $\mathcal{I}^{N-2}_{\rm Hardy}\leq0$. Also, if $|\chi_2\rangle$ is entangled, the violation of
local realism is detected in similar manner.
%\begin{eqnarray}
%% \nonumber to remove numbering (before each equation)
%  p(11|11)\otimes \mathcal{I}^{N-2}_{Hardy}\equiv p(1100|1111)-p(1111|1122)\;\;\;\;\;\;\;\;\nonumber\\-p(1100|1112)-p(1100|1121)\leq0\nonumber\\
%\end{eqnarray}
% Apparently, if both
%$|\chi_1\rangle,|\chi_2\rangle$ are entangled, any of the above
%inequalities is valid for the Bell nonlocality test.

(ii) None of the states $|\chi_1\rangle,|\chi_2\rangle$ is entangled. Up to LU,
one can work with $|\chi_1\rangle=|0\cdots0\rangle$ and
$|\chi_2\rangle=\bigotimes_{i=3}^{N}(f_i|0\rangle+g_i|1\rangle)$,
where $f_i\geq0$.

If the number of coefficients $f_i$ that are equal to zero is
$r\in\{0,1,...,N-3\}$ (without loss of generality we have assumed
that the first $r$ coefficients are zero), then we use the following
Bell inequality
\begin{eqnarray}
% \nonumber to remove numbering (before each equation)
  p(0_1 0_2\cdots0_r|1_1 1_2 \cdots 1_r)\otimes \mathcal{I}^{N-r}_{\rm Hardy} \leq0
\end{eqnarray}
to test the violation of local realism. Let the settings be
\begin{eqnarray}
&&\theta^1_{\mathcal{O}_j}=\phi^1_{\mathcal{O}_j}=0,\;{\rm
for\;}1\leq j\leq
N,\nonumber\\
&&\theta^2_{\mathcal{O}_j}=\vartheta,\;\phi^2_{\mathcal{O}_j}=\frac{\pi}{2},\;{\rm for}\;j=r+1,r+2,\nonumber\\
&&\theta^2_{\mathcal{O}_j}=\pi,\;\phi^2_{\mathcal{O}_j}=0,\;{\rm
for}\;r+3\leq j \leq N,\nonumber
\end{eqnarray}
then quantum mechanics demands
\begin{eqnarray}
% \nonumber to remove numbering (before each equation)
\mathcal{I}_{\rm Hardy}^{\rm
QM}=\frac{\cos^4\frac{\vartheta}{2}}{2^{r+1}}\biggr(
2V\sin\zeta\biggr(\prod_{i=3+r}^{N}
f_i\biggr)\tan^2\frac{\vartheta}{2}-\nonumber\\\biggr[1+\biggr(\prod_{i=3+r}^{N}
f_i\biggr)^2+(1-\biggr(\prod_{i=3+r}^{N}
f_i\biggr)^2)V\cos\zeta\biggr]
 \biggr).\nonumber
\end{eqnarray}
%Note that the first and the second terms in the bracket are both
%positive. For simplicity, we carefully take the value of
%$\theta_{A_2}$ such that the second term is equal to half the first
%term.
Similar to Eq.~(7) in the $N=3$ case in the main text and taking an
appropriate value of $\vartheta$,
%\begin{eqnarray}
%% \nonumber to remove numbering (before each equation)
%\tan^2\frac{\theta_{a_2}}{2}=\frac{(1+\cos^2\phi+(2\nu_1-1)\cos\zeta\sin^2\phi)}{(2\nu_1-1)\sin\zeta\cos\phi},
%\end{eqnarray}
we always have $\mathcal{I}_{\rm Hardy}^{\rm QM}>0$.
%\begin{eqnarray}
%% \nonumber to remove numbering (before each equation)
%\mathcal{I}_{\rm Hardy}^{\rm
%QM}=\frac{\cos^4\frac{\theta_{A_2}}{2}}{2}\left(
%1+\cos^2\phi+(2\nu_1-1)\cos\zeta\sin^2\phi \right)>0.\nonumber
%\end{eqnarray}

If all $f_i=0$ (i.e., $r=N-2$), for an even $N$, the violation of local realism of the
state $\varrho$ is dictated by
\begin{eqnarray}
\mathcal {I}^N=\frac{1}{2^{N-1}}\biggr(\sum_{i_1,i_2,...,i_N=1}^2
Q_{i_1i_2...i_N}\biggr)-Q_{22...2}\leq 1,\label{BI-N}
\end{eqnarray}
where correlation functions $Q_{i_1i_2...i_N}=X_{1i_1}X_{2i_2}\cdots
X_{Ni_N}$, with $X_{ki_k}=\pm 1$, and $i_k$ indicating settings for
the $k$-th party. For $N=2$, the inequality (\ref{BI-N}) reduces to
the CHSH inequality.

The classical bound is obtained as follows. The summation term in
(\ref{BI-N}) is the binomial expansion of $\mathcal
{Y}\equiv\frac{1}{2^{N-1}}(X_{11}+X_{12})(X_{21}+X_{22})\cdots(X_{N1}+X_{N2})$,
which could take only three distinct values $\{-2,0,2\}$. The value
of $Q_{22...2}$ is closely related to $\mathcal {Y}$. Indeed, when
$\mathcal {Y}=2$, then $Q_{22...2}=1$ so that $\mathcal {I}^N=1$;
when $\mathcal {Y}=-2$, then $Q_{22...2}=-1$ so that $\mathcal
{I}^N=-1$; when $\mathcal {Y}=0$, then $\mathcal {I}^N\leq1$ since
$Q_{22...2}$ is no larger than 1. Hence, $\mathcal {I}^N\leq1$ holds
for any local theories.

In quantum mechanics, the correlation function is computed by
$Q_{i_1i_2...i_N}={\rm tr}(\varrho\;\vec{\sigma}_{\vec{n}_{i_1}}
\otimes\vec{\sigma}_{\vec{n}_{i_2}}\otimes\cdots\otimes\vec{\sigma}_{\vec{n}_{i_N}})$,
with
$\vec{n}_{i_k}=(\sin\theta_{ki_k}\cos\phi_{ki_k},\sin\theta_{ki_k}\sin\phi_{ki_k},\cos\theta_{ki_k})$.
We explicitly have
\begin{eqnarray}
Q_{i_1i_2...i_N}&=&\cos\theta_{1i_1}\cos\theta_{2i_2}\cdots\cos\theta_{Ni_N}\nonumber\\
&&+V\sin\zeta\sin\theta_{1i_1}\sin\theta_{2i_2}\cdots
\sin\theta_{Ni_N}\nonumber\\
&&\;\;\;\times\cos(\phi_{1i_1}+\phi_{2i_2}+\cdots+\phi_{Ni_N}).\nonumber
\end{eqnarray}
Taking the settings as
$\phi_{1i_1}=\phi_{2i_2}=\cdots=\phi_{Ni_N}=0$,
$\theta_{N1}=\cdots=\theta_{21}=\pi-\theta_{11}$,
$\theta_{N2}=\cdots=\theta_{22}=\pi-\theta_{12}$, and
$\theta_{12}=\pi$, we get the quantum bound as
\begin{eqnarray}
\mathcal {I}^N_{\rm QM}&=&\frac{1}{2^{N-1}}[2^{N-1}+V\sin\zeta\sin^N\theta_{11}-(1-\cos\theta_{11})^N]\nonumber\\
&=&1+\frac{V\sin\zeta\sin^N\theta_{11}}{2^{N-1}}\biggr(1-\frac{\tan^N\frac{\theta_{11}}{2}}{V\sin\zeta}\biggr).\nonumber
\end{eqnarray}
Clearly, we have $\mathcal {I}^N_{\rm QM}>1$ when
$\tan\frac{\theta_{11}}{2}<(V\sin\zeta)^{1/N}$.

For odd $N$, violation of local realism of the state $\varrho$ is
identified by
\begin{eqnarray}
\mathcal
{I}^N=\frac{1}{2^{N-1}}\biggr(\sum_{i_1,...,i_{N-1}=1}^2\sum_{i_{N}=0}^1
Q_{i_1i_2...i_N}\biggr)-Q_{22...20}\leq 1,\nonumber\\\label{BI-N2}
\end{eqnarray}
where $i_N=0$ indicates that no measurement is performed on the
$N$-th qubit.
%
%The inequality (\ref{BI-N2}) is obtained by fixing $X_{N2}=1$, thus
%the classical bound is also $1$.
Quantum mechanically, we explicitly have
\begin{eqnarray}
Q_{i_1i_2...i_N}%&=&{\rm Tr}(\vec{\sigma}_{\hat{n}_1}
%\otimes\vec{\sigma}_{\hat{n}_2}\otimes\cdots\otimes\vec{\sigma}_{\hat{n}_N}\rho_N)\\
&=&
V\cos\zeta\cos\theta_{1i_1}\cos\theta_{2i_2}\cdots\cos\theta_{Ni_N}\nonumber\\
&&+V\sin\zeta\sin\theta_{1i_1}\sin\theta_{2i_2}\cdots
\sin\theta_{Ni_N}\nonumber\\
&&\;\;\;\times\cos(\phi_{1i_1}+\phi_{2i_2}+\cdots+\phi_{Ni_N}),\nonumber
\end{eqnarray}
and, $Q_{i_1i_2...i_{N-1}0}=\cos\theta_{1i_1}\cos\theta_{2i_2}\cdots\cos\theta_{Ni_N}$.
%\begin{eqnarray}
%Q_{i_1i_2...i_{N-1}0}%&=&{\rm
%%Tr}(\vec{\sigma}_{\hat{n}_1}
%%\otimes\vec{\sigma}_{\hat{n}_2}\otimes\cdots\otimes\vec{\sigma}_{\hat{n}_{N-1}}\otimes
%%\textbf{1 }\;\rho_N)\\
%&=&\cos\theta_{1i_1}\cos\theta_{2i_2}\cdots\cos\theta_{Ni_N}.
%\nonumber
%\end{eqnarray}
By taking settings as
$\phi_{1i_1}=\phi_{2i_2}=\cdots=\phi_{Ni_N}=0$,
$\theta_{ni_n}=\frac{\pi}{2}\times i_n,\;(n\neq1)$, and
$\theta_{12}=0$, all correlation functions vanish except the
following ones: $Q_{111...11}=V\sin\zeta\sin\theta_{11}$,
$Q_{122...20}=-\cos\theta_{11}$, and $Q_{222...20}=-1$. Hence, the
quantum prediction of $\mathcal {I}^N$ in (\ref{BI-N2}) is
\begin{eqnarray}
\mathcal {I}_N^{\rm QM}
&=&1+\frac{\cos^2\frac{\theta_{11}}{2}}{2^{N-2}}(V\sin\zeta\tan\frac{\theta_{11}}{2}-1).\nonumber
\end{eqnarray}
Clearly, we have $\mathcal {I}_N^{\rm QM}>1$ when
$\tan\frac{\theta_{11}}{2}>\frac{1}{V\sin\zeta}$, proving
Theorem 2.  \hfill\endproof

{\bf Acknowledgements}

This work is supported by the National Basic Research Program
(973 Program) of China (Grant Nos.  2012CB921900, 2011CBA00200, and
2011CB921200) and and the NSF of China (Grant Nos. 11175089, 11475089, 11275182, and 60921091). J.L.C. and L.C.K. are partly supported by the
National Research Foundation and the Ministry of
Education, Singapore. MZ is supported by an FNP TEAM project and an MNiSW project Ideas-Plus (Wydajnosc kwantowo-mechanicznych zasobow).

{\bf Author Contributions}

J.L.C. initiated the idea. J.L.C., H.Y.S., Z.P. X, Y.C.W and C.W.
established the proof of the theorems. J.L.C., H.Y.S., C.W., M.Z.,
and L.C.K. wrote the main manuscript text. X.J.Y. prepared figure.
All authors reviewed the manuscript.

%{\bf Supplementary Information} is linked to the online version of
%the paper at www.nature.com/nature.

{\bf Additional Information}

\textbf{Competing financial interests:} The authors declare no
competing financial interests.

\textbf{Reprints and permission} information is available at
www.nature.com/reprints.

\newpage

\begin{figure}[t]
\includegraphics[width=90mm]{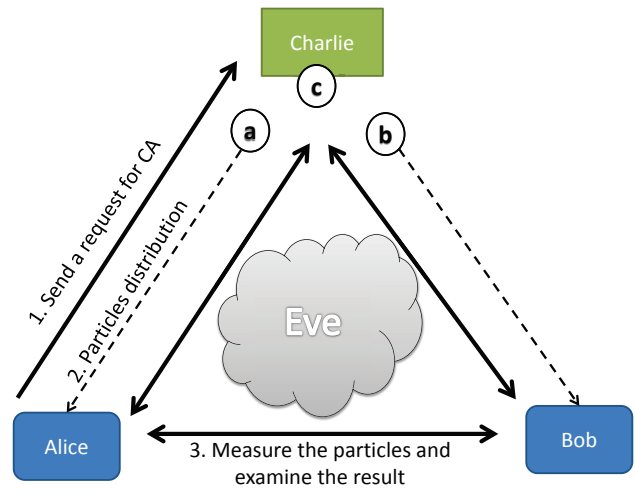}\\
\caption{(Color online) The illustration of quantum certificate authorization protocol. Upon Alice's request for identification of Bob, Charlie produces a three-qubit state $\varrho$ and then distributes the first qubit (represented as a ball labeled by ``a'', similarly for the others) to Alice , the second qubit ``b" to Bob, and keeps the third qubit ``c". To ensure the security, Charlie randomly measures his qubit along $|\chi_1\rangle$ or $|\chi_2\rangle$, Alice and Bob randomly measure their qubits along $\vec{z}$ or $\vec{x}$. Such a measurement can be repeated for large enough times. Finally Charlie performs a random inspection
 %in three steps (see Remark 2 for detail)
 to see whether Alice and Bob are able to share secret keys at the quantum level that defy Eve's eavesdrop.
%Charlie randomly picks measuring directions from bases $\{\hat{x}_A\otimes\hat{x}_B, \hat{z}_A\otimes\hat{z}_B\}$ and asks Alice and Bob to measure their qubits accordingly. He then measure his qubit ``c" along $|\chi_1\rangle$ if Alice's and Bob's results along $\hat{z}$ are both $0$, or along $|\chi_2\rangle$ if their results are both $1$. Finally Charlie takes a random inspection to see whether the remaining parties can be steered into two pure states when Alice and Bob measure along $\hat{z}_A\otimes\hat{z}_B$. If this is true, they further check whether the whole state they share is entangled by comparing the results of $\hat{x}_A\otimes\hat{x}_B$. If this is also true, then it is concluded that Alice and Bob share secret keys at the quantum level that defy Eve's eavesdrop.
}\label{fig2}
\end{figure}

\end{document}